# Structural, Elastic, Electronic and Optical Properties of a New Layered-Ternary $Ta_4SiC_3$ Compound

M. S. Islam and A.K.M.A. Islam[*]

*Department of Physics, Rajshahi University, Rajshahi-6205, Bangladesh*

**ABSTRACT**

We propose a new layered-ternary $Ta_4SiC_3$ with two different stacking sequences ($\alpha$- and $\beta$-phases) of the metal atoms along $c$ axis and study their structural stability. The mechanical, electronic and optical properties are then calculated and compared with those of other compounds $M_4AX_3$ (M = V, Nb, Ta; A = Al, Si and X = C). The predicted compound in the $\alpha$-phase is found to possess higher hardness than any of these compounds. The independent elastic constants of the two phases are also evaluated and the results discussed. The electronic band structures for $\alpha$- and $\beta$-$Ta_4SiC_3$ show metallic conductivity. Ta $5d$ electrons are mainly contributing to the total density of states (DOS). We see that the hybridization peak of Ta $5d$ and C $2p$ lies lower in energy and the Ta $5d$-C $2p$ bond is stronger than Ta $5d$-Si $3p$ bond. Further an analysis of the different optical properties shows the compound to possess improved behavior compared to similar types of compounds.

## 1. Introduction

The layered-ternary compounds in the so-called MAX phases exist as $M_2AX$ (211), $M_3AX_2$ (312) and $M_4AX_3$ (413) compounds, where M is an early transition metal, A is an A-group element (usually belonging to the groups IIIA and IVA) and X is either C or N. All these compounds show interesting properties, e.g. electrical and thermal conduction, high bulk modulus, high melting point, low density, low Vickers hardness, damage-tolerance, thermal shock resistance and so on, and have been investigated extensively [1-20].

Currently there are eight $M_4AX_3$ (413) compounds, such as $Ti_4AlN_3$, $Ti_4SiC_3$, $Ti_4GeC_3$, $Ti_4GaC_3$, $Ta_4AlC_3$, $Nb_4AlC_3$, $V_4AlC_3$, and $V_4AlC_{2.69}$ which have been synthesized [9, 12-19]. Among these $Ta_4AlC_3$, a relatively new member of 413 phases was synthesized in the form of polycrystals [8-11] and single-crystals [12]. Subsequently two new 413 phases, $Nb_4AlC_3$ [16] and $V_4AlC_3$ [17] were synthesized and characterized. But $Ta_4AlC_3$ has drawn a lot of attention recently [8-13]. The bulk modulus of $Ta_4AlC_3$ is determined to be 261 GPa, which is so far the highest among all the MAX phases reported [7]. Further it has been found to possess high flexural strength, high fracture toughness with good electrical and thermal and improved mechanical properties [13].

The polymorphs of $Ta_4AlC_3$ have been identified by several workers [7, 10-12]. The two phases of $Ta_4AlC_3$ crystallize in a hexagonal structure with the space group $P6_3/mmc$ but have different atomic positions. One phase is $\alpha$-$Ta_4AlC_3$, which is the same as $Ti_4AlN_3$, stacking sequence of the metal atom Ta and Al in the order of AB<u>A</u>BAC<u>B</u>CBC along *c* axis, reported by Etzkorn *et al*. [12]. The other phase, reported by Lin *et al*. [11, 20], is the so-called *β*-Ta$_4$AlC$_3$ which stacks in ABABABABAB along *c* axis.

---

[*] Corresponding author:
*E-mail address*: azi46@ru.ac.bd (A.K.M.A. Islam)

As mentioned earlier Ta$_4$AlC$_3$ has excellent properties and it is envisaged that there may be other compounds that could be predicted with a different A-group element which could show more desirable properties. In order to explore new field of 413 phases we propose a new layered-ternary Ta$_4$SiC$_3$ and study its stability. Elastic properties are essential to the understanding of the macroscopic mechanical properties of crystals because they are related to various fundamental solid state properties and thermodynamic properties. A first-principles study of structural, elastic (for both mono- and poly-crystalline aggregate) and electronic properties in the framework of density functional theory would be made. Further the optical properties (dielectric function, absorption spectrum, conductivity, energy-loss spectrum and reflectivity) will be calculated and discussed.

## 2. Computational methods

The first-principles *ab-initio* calculations are performed using the CASTEP code [21] in the framework of density functional theory (DFT) with generalized gradient approximation (GGA) and default the Perdew-Burke-Ernzerhof (PBE) [22]. The interactions between ion and electron are represented by ultrasoft Vanderbilt-type pseudopotentials for Ta, Si and C atoms [23]. The elastic constants are calculated by the 'stress–strain' method. All the calculating properties for Ta$_4$SiC$_3$ used a plane-wave cutoff energy 450 eV and 9×9×2 Monkhorst-Pack [24] grid for the sampling of the Brillouin zone. Geometry optimization is conducted using convergence thresholds of 5×10$^{-6}$ eV atom$^{-1}$ for the total energy, 0.01 eVÅ$^{-1}$ for the maximum force, 0.02 GPa for maximum stress and 5×10$^{-4}$ Å for maximum displacement.

## 3. Results and discussions

### 3.1 Structural properties

The proposed compound Ta$_4$SiC$_3$ is first assumed to have a crystal structure similar to Ta$_4$AlC$_3$ and also other M$_4$AX$_3$ compounds. We know that Ta$_4$AlC$_3$ possesses two types of structures, i.e. *α*- and *β*- forms with two different stacking sequences. We then perform the geometry optimization as a function of the normal stress by minimizing the total energy of the proposed compound for both these phases, i.e. $\alpha$- and $\beta$-Ta$_4$SiC$_3$. The calculated total energy (-2247.77 eV) of $\alpha$-Ta$_4$SiC$_3$ is more than that (-2246.74 eV) of $\beta$-Ta$_4$SiC$_3$, indicating $\alpha$-Ta$_4$SiC$_3$ to be energetically more favourable. The crystal structures of $\alpha$-Ta$_4$SiC$_3$ and $\beta$-Ta$_4$SiC$_3$ are illustrated in Fig. 1. The optimized parameters for both the phases are shown in (Table 1), along with those of other M$_4$AX$_3$, (M=V, Nb, Ta) compounds.

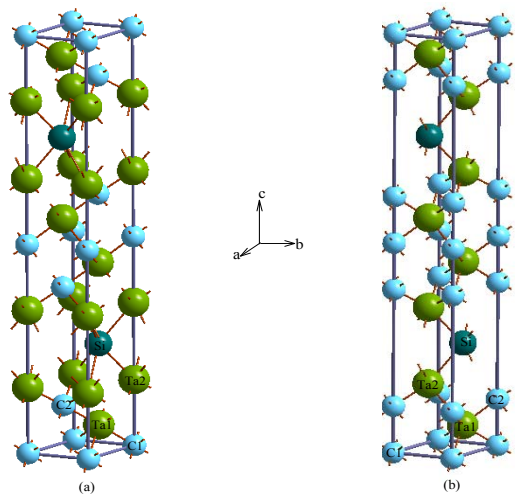

Fig. 1. Crystal structures of (a) *α*-Ta$_4$SiC$_3$ and (b) *β*-Ta$_4$SiC$_3$.

Table 1. The optimized structural parameters for both the structures of Ta$_4$SiC$_3$.

|   | $a$ (Å) | $c$ (Å) | $c/a$ | $V_0$ (Å$^3$) | Ref. |
|---|---|---|---|---|---|
| $\alpha$-Ta$_4$SiC$_3$ | 3.2514 | 22.8900 | 7.04 | 209.60 | This |
| $\beta$-Ta$_4$SiC$_3$ | 3.1943 | 23.7290 | 7.43 | 209.70 | This |
| $\alpha$-Ta$_4$AlC$_3$ | 3.134 | 24.268 | 7.74 | 206.42 | [25]$^T$ |
|   | 3.138 | 24.163 | 7.70 | 206.05 | [26]$^T$ |
|   | 3.312 | 24.111 | 7.75 | 229.04 | [12]$^E$ |
|   | 3.109 | 24.708 | 7.74 | 206.82 | [10]$^E$ |
| $\beta$-Ta$_4$AlC$_3$ | 3.070 | 24.198 | 7.88 | 197.50 | [27]$^T$ |
|   | 3.095 | 24.714 | 7.98 | 205.01 | [26]$^T$ |
|   | 3.087 | 23.700 | 7.68 | 195.60 | [7]$^E$ |
|   | 3.309 | 23.078 | 7.67 | 218.83 | [27]$^E$ |

T= Theoretical, E = Experimental.

### 3.2 Elastic constants and mechanical stability

To study the mechanical properties of $\alpha$- and $\beta$-Ta$_4$SiC$_3$, we have calculated the elastic constants $C_{ij}$, bulk modulus $B$, shear modulus $G$, Young's modulus $E$, Poisson's ratio $v$. The calculated results are illustrated in Table 2 along with other available theoretical results of several M$_4$AX$_3$ compounds. Both the structures have six elastic constants ($C_{11}$, $C_{12}$, $C_{13}$, $C_{33}$, $C_{44}$ and $C_{66}$) and only five of them are independent, since $C_{66} = (C_{11} - C_{12})/2$. Both $\alpha$- and $\beta$-Ta$_4$SiC$_3$ have $C_{11}$ values smaller than those of $\alpha$- and $\beta$-Ta$_4$AlC$_3$, which leads to the lower resistances against the principal strain $\varepsilon_{11}$. The $C_{12}$ of $\alpha$-Ta$_4$SiC$_3$ is 33 GPa larger than that of $\alpha$-Ta$_4$AlC$_3$ and this leads to higher resistances against the principal strain $\varepsilon_{12}$. The $C_{44}$ of $\alpha$-Ta$_4$SiC$_3$ is 74 GPa larger than that of $\beta$-Ta$_4$SiC$_3$, thereby showing higher resistances to basal and prismatic shear deformations.

The criteria of mechanical stability are essential to illustrate a stable compound. The Born stability criteria [31] for $\alpha$- and $\beta$-Ta$_4$SiC$_3$ are as follows:

$$C_{11} > 0, \quad C_{11} - C_{12} > 0,$$
$$C_{44} > 0, \quad (C_{11} + C_{12}) C_{33} - 2C^2_{13} > 0. \tag{1}$$

From the calculated elastic constants it is easy to see that these criteria are satisfied for both $\alpha$-Ta$_4$SiC$_3$ and $\beta$-Ta$_4$SiC$_3$ and hence they are stable at zero pressure.

Table 2. Elastic constants $C_{ij}$, the bulk modulus $B$, shear modulus $G$, Young's modulus $E$ (all in GPa), Poisson's ratio $\nu$, anisotropic factor $A$, linear compressibility ratio $k_c/k_a$ and ratio $G/B$ at zero pressure.

|  | Monocrystal | | | | | | Polycrystal | | | | | | |
| --- | --- | --- | --- | --- | --- | --- | --- | --- | --- | --- | --- | --- | --- |
|  | $C_{11}$ | $C_{12}$ | $C_{13}$ | $C_{33}$ | $C_{44}$ | $C_{66}$ | $B$ | $G$ | $E$ | $\nu$ | $A$ | $k_c/k_a$ | $G/B$ |
| $\alpha$-Ta$_4$SiC$_3$[a] | 396 | 190 | 180 | 391 | 207 | 103 | 254 | 138 | 350 | 0.27 | 2.00 | 1.07 | 0.54 |
| $\beta$-Ta$_4$SiC$_3$[a] | 397 | 148 | 190 | 397 | 133 | 124 | 250 | 121 | 312 | 0.29 | 1.06 | 0.80 | 0.48 |
| $\alpha$-Ta$_4$AlC$_3$[b] | 454 | 157 | 156 | 376 | 201 | 149 | 247, 239[e] | 161 | 397 | 0.23 |  |  | 0.65 |
| $\beta$-Ta$_4$AlC$_3$[b] | 452 | 152 | 150 | 441 | 145 | 150 | 250, 240[e] | 148 | 370 | 0.25 |  |  | 0.59 |
| $\alpha$-Nb$_4$SiC$_3$[c] | 403 | 167 | 165 | 374 | 195 |  | 241 | 142 |  |  |  |  | 0.59 |
| $\alpha$-Nb$_4$AlC$_3$[c] | 413 | 124 | 135 | 328 | 161 |  | 214 | 144 |  |  |  |  | 0.67 |
| V$_4$AlC$_3$[d] | 458 | 107 | 110 | 396 | 175 |  | 218 | 170 |  |  |  | 1.21 | 0.78 |

[a]This work, [b]Ref. [26], [c]Ref. [36], [d]Ref. [37], [e]fitted value.

The theoretical polycrystalline elastic moduli for $\alpha$- and $\beta$-Ta$_4$SiC$_3$ may be computed from the set of independent elastic constants. Hill's [28] proved that the Voigt and Reuss equations represent upper and lower limits of the true polycrystalline constants. According to Hill's observation, the polycrystalline moduli are defined as the average values of the Voigt ($B_V$, $G_V$) and Reuss ($B_R$, $G_R$) moduli. From Hill's observation, the value of bulk modulus (in GPa) $B = (B_V + B_R)/2 = B_H$ (Hill's bulk modulus), where $B_V$ is the Voigt's bulk modulus and $B_R$ is the Reuss's bulk modulus. The value of shear modulus (in GPa) $G = (G_V + G_R)/2 = G_H$ (Hill's shear modulus), where $G_V$ is the Voigt's shear modulus and $G_R$ is the Reuss's shear modulus. These expressions for Reuss and Voigt moduli can be found in ref. [29]. The polycrystalline Young's modulus $E$ (in GPa) and the Poisson's ratio $\nu$ are then calculated using the relationships [30]: $E = 9BG/(3B + G)$ and $\nu = (3B - E)/6B$ respectively. The bulk modulus $B$ of $\alpha$-Ta$_4$SiC$_3$ is 7 GPa higher than that of $\alpha$-Ta$_4$AlC$_3$. Table 2 shows that $\alpha$-Ta$_4$SiC$_3$ possesses higher hardness due to its higher bulk modulus than the other M$_4$AX$_3$ compounds.

Using the calculated elastic constants $C_{ij}$, there are different ways to represent the elastic anisotropy of crystals. For this purpose, so-called shear anisotropy ratio $A = 2C_{44}/(C_{11} - C_{12})$ is often used [32]. The factor $A = 1$ represents complete isotropy, while values smaller or greater than this measure the degree of anisotropy. Therefore, $\alpha$-Ta$_4$SiC$_3$ shows completely anisotropic behavior but $\beta$-Ta$_4$SiC$_3$ is nearly isotropic (Table 2).

The parameter $k_c/k_a = (C_{11} + C_{12} - 2C_{13})/(C_{33} - C_{13})$ is used, which expresses the ratio between linear compressibility coefficients of hexagonal crystals [33]. The data obtained $k_c/k_a = 1.07$ demonstrate that the compressibility for $\alpha$-Ta$_4$SiC$_3$ along c axis is greater than along a axis and $k_c/k_a = 0.8$ that the compressibility for $\beta$-Ta$_4$SiC$_3$ along $c$ axis is smaller than along a axis.

According to Pugh's criteria [34], a material should behave in a ductile manner, if $G/B < 0.5$, otherwise it should be brittle. For $\alpha$-Ta$_4$SiC$_3$, $G/B = 0.54$, i.e. $\alpha$-Ta$_4$SiC$_3$ is slightly above the borderline and for $\beta$-Ta$_4$SiC$_3$, $G/B = 0.48$, i.e. $\beta$-Ta$_4$SiC$_3$ will behave as a ductile material. For ductile metallic materials, Poisson's ratio $\nu$ is typically 0.33 [35], $\beta$-Ta$_4$SiC$_3$ shows greater ductility than that of $\alpha$-Ta$_4$SiC$_3$.

The Debye temperature $\Theta_D$ is proportional to the average elastic wave velocity $v_a$. Then the Debye temperature $\Theta_D$ may be estimated from the average elastic wave velocity $v_a$ [38]:

$$\Theta_D = \frac{h}{k_B}\left(\frac{3n}{4\pi V_0}\right)^{\frac{1}{3}} v_a \qquad (2)$$

where $h$ is Planck's constant, $k_B$ is the Boltzmann's constant, $V_0$ is the volume of unit cell and $n$ is the number of atoms in unit cell. Now the average elastic wave velocity $v_a$ (m/sec) can be obtained from the transverse $v_t$ and longitudinal wave velocity $v_l$, respectively (see [29]). Table 3 shows that $\alpha$-Ta$_4$SiC$_3$ possesses larger elastic wave velocities and larger Debye temperatures $\Theta_D$. The Debye temperatures for the predicted compounds in $\alpha$- and $\beta$-phases are found to be lower than those of Ti$_3$SiC$_2$ (780 K), Ti$_4$AlN$_3$ (762 K) and Ti$_3$AlC$_2$ (758 K) [39].

Table 3. The transverse, longitudinal, average elastic wave velocities ($v_t$, $v_l$, $v_a$ in m/sec) and Debye temperature ($\Theta_D$ in K) for $\alpha$-Ta$_4$SiC$_3$ and $\beta$-Ta$_4$SiC$_3$.

| Phase | $\rho$ | $v_t$ | $v_l$ | $v_a$ | $\Theta_D$ |
|---|---|---|---|---|---|
| $\alpha$-Ta$_4$SiC$_3$ | 12.48 | 3325 | 5924 | 4360 | 535 |
| $\beta$-Ta$_4$SiC$_3$ | 12.46 | 3116 | 5745 | 4092 | 502 |

### 3.3 Electronic properties

The energy band structures and DOS of $\alpha$- and $\beta$-Ta$_4$SiC$_3$ are illustrated in Figs. 2 and 3. The valence and conduction bands overlap considerably and there are many bands crossing the Fermi level. Thus α- and β-Ta$_4$SiC$_3$ should show metallic conductivity. These energy bands around the Fermi energy are mainly from the Ta 5$d$ states, suggesting that the Ta 5$d$ states dominate the conductivity of Ta$_4$SiC$_3$. It is apparent that Si 3$s$/3$p$ electrons do not contribute significantly at the Fermi level for a scooping effect due to the presence of the Ta 5$d$ states. The lowest valence bands from -14 to -10 eV below Fermi level ($E_F$ = 0 eV) arise mainly from the C 2$s$ states, with small contribution from the Ta 5$d$ states. The higher valence bands between -8 and 0 eV are dominated by hybridizing Ta 5$d$, Si 3$s$/3$p$ and C 2$p$ states. Similar results are found with previous reports on M$_4$AX$_3$ phaes [30, 36, 37]. As is evident the hybridization peak of Ta 5$d$ and C 2$p$ lies lower in energy than that of Ta 5$d$ and Si 3$p$ states. This suggests that the Ta 5$d$-C 2$p$ bond is stronger than the Ta 5$d$-Si 3$p$ bond.

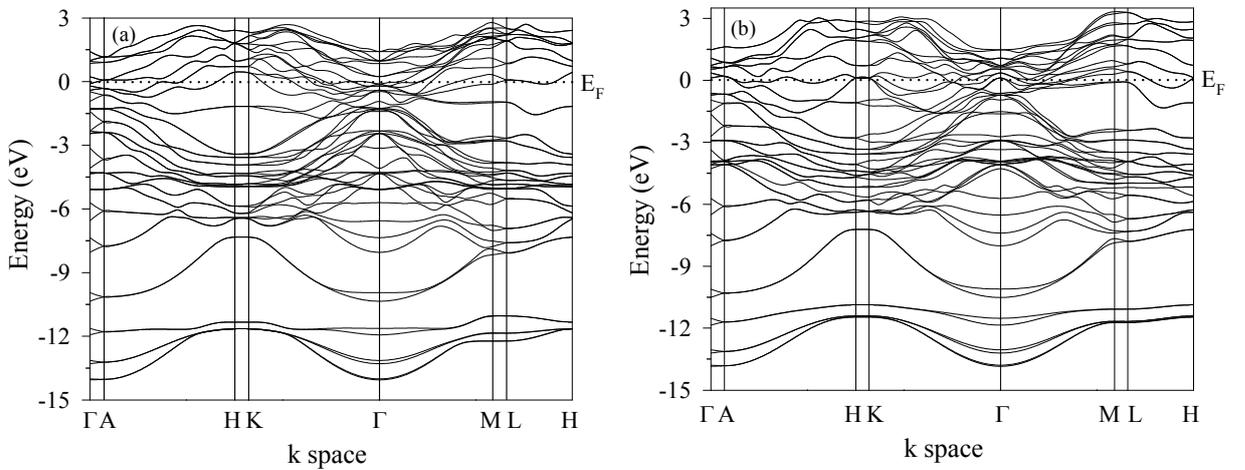

Fig. 2. Band structures of (a) $\alpha$-Ta$_4$SiC$_3$ and (b) $\beta$-Ta$_4$SiC$_3$.

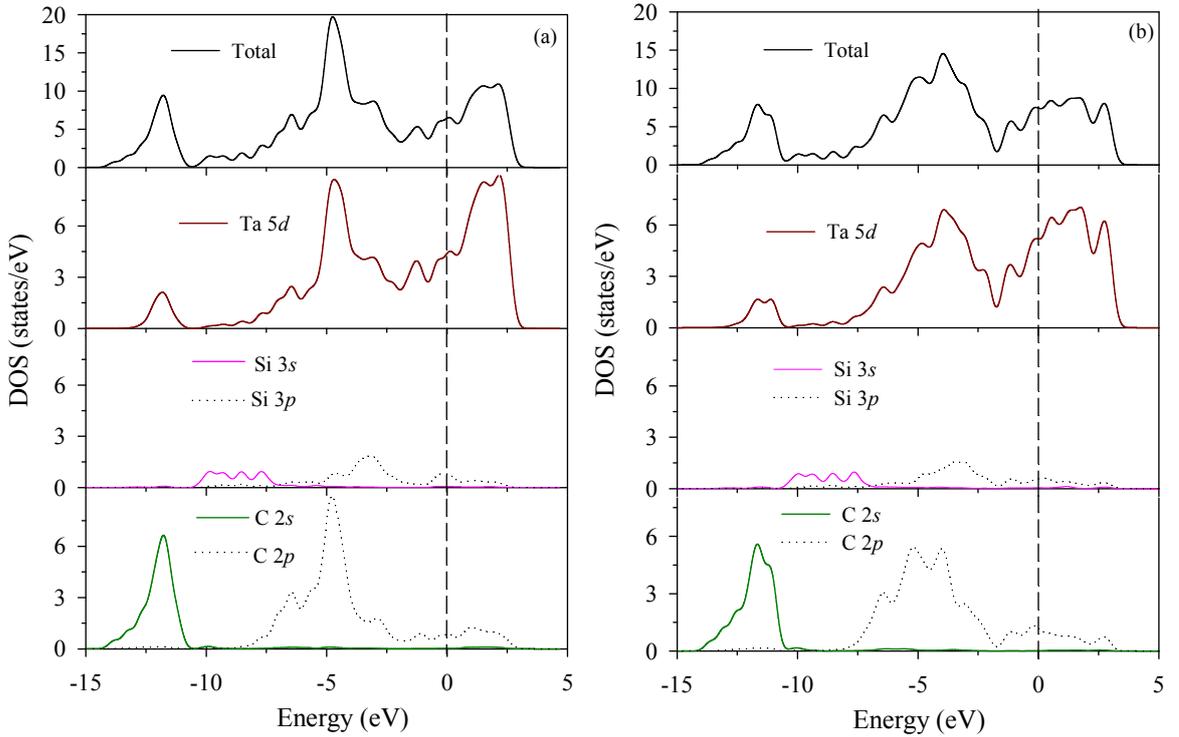

Fig. 3. Total and partial DOSs of (a) α-Ta$_4$SiC$_3$ and (b) β-Ta$_4$SiC$_3$.

## 3.4 Optical properties

The optical constants of α- and β-Ta$_4$SiC$_3$: real part and imaginary part of dielectric function $\varepsilon_1(\omega)$ and $\varepsilon_2(\omega)$, absorption, photoconductivity, energy-loss function, refractive index and reflectivity spectrum are shown in Figs. 4 and 5. We used a 0.5 eV Gaussian smearing for all calculations. For the imaginary part $\varepsilon_2(\omega)$ of the dielectric function, the peak around 1 eV is due to transitions within the Ta 5$d$ bands and the $\varepsilon_2(\omega)$ spectrum above 5 eV arises from Si/C $p \rightarrow$ Ta $d$ electronic transitions. The large negative value of $\varepsilon_1(\omega)$ indicates that the Ta$_4$SiC$_3$ crystal has a Drude-like behavior.

In Fig. 4b, we show the absorption spectrum with one peak between 7.5 to 9 eV for α-Ta$_4$SiC$_3$ which rises and then decreases rapidly in the high-energy region. Nearly same feature can be seen for β-Ta$_4$SiC$_3$ but with two small peaks in the low energy side. The peak at around 7.4-8 eV is associated with the transition from Si/C $p$ to Ta $d$ states.

For both the structures, photoconductivity starts with zero photon energy. This shows that the material has no band gap. Moreover, the photoconductivity and hence electrical conductivity of a material increases as a result of absorbing photons [40]. The photoconductivity of $\alpha$-$Ta_4SiC_3$ (Fig. 4c) shows three peaks at 1.8, 5.6, 14.4 (small) eV. $\beta$-$Ta_4SiC_3$ also shows peaks but at 1.14, 3.5, and 5.7 eV.

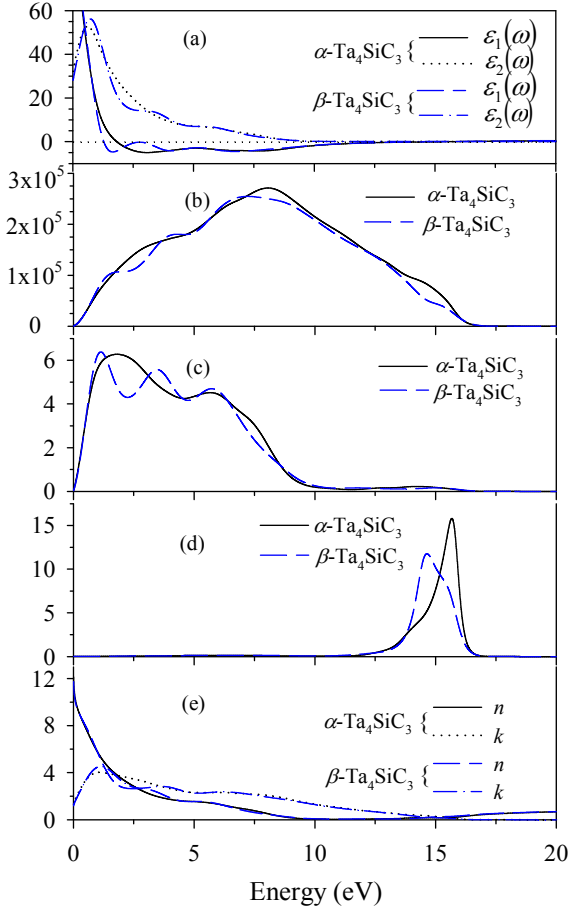

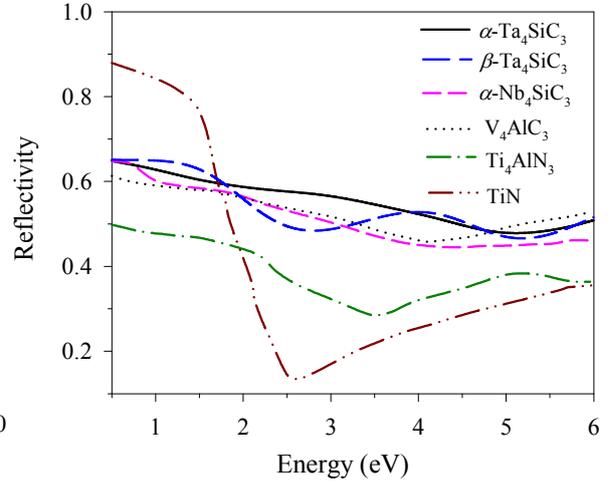

Fig. 4. Optical constants of $\alpha$- and $\beta$-$Ta_4SiC_3$: (a) Imaginary part $\varepsilon_2(\omega)$ and real part $\varepsilon_1(\omega)$ of the dielectric function $\varepsilon(\omega)$, (b) absorption spectrum, (c) photo-conductivity, (d) energy-loss spectrum, and (e) refractive index.

Fig. 5. Reflectivity spectra of $\alpha$-$Ta_4SiC_3$, $\alpha$-$Nb_4SiC_3$, $V_4AlC_3$, $Ti_4AlN_3$ and TiN.

The energy loss of a fast electron traversing in the material is manifested in the energy-loss spectrum [41]. Its peak is defined as the bulk plasma frequency $\omega_p$, which occurs at $\varepsilon_2 < 1$ and $\varepsilon_1$ reaches the zero point [42, 43]. In the energy-loss spectrum (Fig. 4d), we see that the plasma frequency $\omega_p$ for $\alpha$-$Ta_4SiC_3$ is 15.7 eV (14.6 for $\beta$-phase). When the frequency of incident light is higher than the plasma frequency, the material becomes transparent. The real part (refractive index, $n$) and imaginary part (extinction coefficient, $k$) of the complex refractive index have been shown in Fig. 4e.

The reflectivity as a function of energy for both phases $Ta_4SiC_3$ is illustrated in Fig. 5. The spectra of $V_4AlC_3$ [37], $\alpha$-$Nb_4SiC_3$, $Ti_4AlN_3$ and TiN [36] are also shown in the figure for comparison. The reflectance spectrum of $\alpha$-$Ta_4SiC_3$ is rather flat with no strong edge and color (but $\beta$-$Ta_4SiC_3$ has a reflectance drop at ~2.77 eV). These characteristics of $\alpha$-$Ta_4SiC_3$ are nearly similar to those of $\alpha$-$Nb_4SiC_3$, $V_4AlC_3$ (see Fig. 5). On the other hand, the spectrum of TiN has sharp reflectance drop between 1.35 and 2.6 eV, which is characteristic of high conductance. TiN has high reflectivity in the infrared and low reflectivity (transparency) for shorter wavelengths [44]. The low reflectance in the region of blue and violet light (2.8 - 3.5 eV) for TiN gives its goldlike color [45]. Therefore, the spectrum of TiN is selective. We can see that the spectrum of $\alpha$-Ta$_4$SiC$_3$ is nonselective, similar to those of transition metals. According to the nonselective characteristic of the reflectance spectrum, $\alpha$-Ta$_4$SiC$_3$ could reduce solar heating and enhance the infrared emittance and therefore the equilibrium temperature of its surface will be moderate in strong sunlight.

In Fig. 5, the reflectance spectra for $\alpha$-Ta$_4$SiC$_3$ and $\alpha$-Nb$_4$SiC$_3$ at 0.5 eV are the same but at 6.0 eV, the reflectivity of $\alpha$-Ta$_4$SiC$_3$ is higher than that of $\alpha$-Nb$_4$SiC$_3$. This indicates that the reflectivity of $\alpha$-Ta$_4$SiC$_3$ is always higher than those of $\alpha$-Nb$_4$SiC$_3$, V$_4$AlC$_3$ Ti$_4$AlN$_3$. Therefore, the capability of $\alpha$-Ta$_4$SiC$_3$ to reflect solar light is stronger than $\alpha$-Nb$_4$SiC$_3$, V$_4$AlC$_3$ and Ti$_4$AlN$_3$.

## 4. Conclusions

In conclusion, a new layered-ternary Ta$_4$SiC$_3$ compound in two phases has been predicted using first-principles calculations. The structural stability and mechanical, electronic and optical properties are then made. The elastic constants, bulk modulus, shear modulus and Young's moduli of $\alpha$- and $\beta$-Ta$_4$SiC$_3$ are compared to those of other similar M$_4$AX$_3$ compounds. Both the structures are found to be stable mechanically. The results show that both the phases show ductile behavior but $\alpha$- phase is found to possess higher hardness than any of the similar M$_4$AX$_3$ compounds. Further, $\alpha$-Ta$_4$SiC$_3$ shows largely anisotropic elasticity but $\beta$-Ta$_4$SiC$_3$ is nearly isotropic. The electronic band structures for both the phases show metallic conductivity. Moreover, the Ta-Si bonding is weaker than the Ta-C bonding in Ta$_4$SiC$_3$ indicating that the Ta-C bond is more resistant to deformation than the Ta-Si bond. Lastly, the optical constants e.g. the reflectivity spectrum of Ta$_4$SiC$_3$ indicates that the predicted compound might be a better candidate material as a coating to avoid solar heating than the other existing $\alpha$-Nb$_4$SiC$_3$, V$_4$AlC$_3$ Ti$_4$AlN$_3$ and TiN compounds. It is expected that our prediction would stimulate experimental study on the compound.